\documentstyle[12pt]{report}
\newskip\humongous \humongous=0pt plus 1000pt minus 1000pt

\newif\ifdtup

\def\beq{\begin{equation}}
\def\eeq{\end{equation}}

\def\beqn{\begin{eqnarray}}
\def\eeqn{\end{eqnarray}}
\relax

\def\dotx{\dotx{\dot\overline{x}}}

\relax
\catcode`\@=11

\@addtoreset{equation}{section}
\def\theequation{\arabic{section}.\arabic{equation}}

\def\@normalsize{\@setsize\normalsize{15pt}\xiipt\@xiipt
\abovedisplayskip 14pt plus3pt minus3pt%
\belowdisplayskip \abovedisplayskip
\abovedisplayshortskip  \z@ plus3pt%
\belowdisplayshortskip  7pt plus3.5pt minus0pt}

\def\small{\@setsize\small{13.6pt}\xipt\@xipt
\abovedisplayskip 13pt plus3pt minus3pt%
\belowdisplayskip \abovedisplayskip
\abovedisplayshortskip  \z@ plus3pt%
\belowdisplayshortskip  7pt plus3.5pt minus0pt
\def\@listi{\parsep 4.5pt plus 2pt minus 1pt
            \itemsep \parsep
            \topsep 9pt plus 3pt minus 3pt}}

\@twosidetrue

\relax

\catcode`@=12

\catcode`\@=11

\def\section{\@startsection{section}{1}{\z@}{3.5ex plus 1ex minus
   .2ex}{2.3ex plus .2ex}{\large\bf}}

\def\thesection{\Roman{section}.}

\def\appendix{\setcounter{section}{0}
	\def\thesection{APPENDIX \Alph{section}:}
	\def\theequation{\Alph{section}.\arabic{equation}}}

\def\FERMIPUB{}
\def\FERMILABPub#1{\def\FERMIPUB{#1}}
\def\ps@headings{\def\@oddfoot{}\def\@evenfoot{}
\def\@oddhead{\hbox{}\hfill
	\makebox[.5\textwidth]{\raggedright\ignorespaces --\thepage{}--
	\hfill {\rm FERMILAB--Pub--\FERMIPUB}}}
\def\@evenhead{\@oddhead}
\def\subsectionmark##1{\markboth{##1}{}}
}

\ps@headings

\catcode`\@=12

\relax

\def\figcap{\section*{Figure Captions\markboth
	{FIGURECAPTIONS}{FIGURECAPTIONS}}\list
	{Fig. \arabic{enumi}:\hfill}{\settowidth\labelwidth{Fig. 999:}
	\leftmargin\labelwidth
	\advance\leftmargin\labelsep\usecounter{enumi}}}
 \relax
\def\tablecap{\section*{Table Captions\markboth
	{TABLECAPTIONS}{TABLECAPTIONS}}\list
	{Table \arabic{enumi}:\hfill}{\settowidth\labelwidth{Table 999:}
	\leftmargin\labelwidth
	\advance\leftmargin\labelsep\usecounter{enumi}}}
 \relax
\def\reflist{\section*{References\markboth{REFLIST}{REFLIST}}\list
	{[\arabic{enumi}]\hfill}{\settowidth\labelwidth{[999]}
	\leftmargin\labelwidth
	\advance\leftmargin\labelsep\usecounter{enumi}}}
 \relax

\catcode`\@=11

\def\FERMIPUB{}
\def\FERMILABPub#1{\def\FERMIPUB{#1}}
\def\ps@headings{\def\@oddfoot{}\def\@evenfoot{}
\def\@oddhead{\hbox{}\hfill
	\makebox[.5\textwidth]{\raggedright\ignorespaces --\thepage{}--
	\hfill {\rm FERMILAB--Pub--\FERMIPUB}}}
\def\@evenhead{\@oddhead}
\def\subsectionmark##1{\markboth{##1}{}}
}

\ps@headings

\relax
\catcode`\@=11

\def\FERMIPUB{}
\def\FERMILABPub#1{\def\FERMIPUB{#1}}
\def\ps@headings{\def\@oddfoot{}\def\@evenfoot{}
\def\@oddhead{\hbox{}\hfill
	\makebox[.5\textwidth]{\raggedright\ignorespaces
--\thepage{}--
	\hfill {\rm FERMILAB--Pub--\FERMIPUB}}}
\def\@evenhead{\@oddhead}
\def\subsectionmark##1{\markboth{##1}{}}
}

\ps@headings

\catcode`\@=12

\relax

\catcode`\@=11

\def\FERMIPUB{}
\def\FERMILABPub#1{\global\def\FERMIPUB{#1}}
\def\ps@headings{\def\@oddfoot{\hfill \thepage{} \hfill}
\def\@evenfoot{\hfill \thepage{} \hfill}
\def\@oddhead{\hbox{}\hfill
	\makebox[.5\textwidth]{\raggedright\ignorespaces
	\hfill {\rm FERMILAB--Pub--\FERMIPUB}}}
\def\@evenhead{\@oddhead}
\def\subsectionmark##1{\markboth{##1}{}}
}

\ps@headings

\catcode`\@=12

\relax

\renewcommand{\thefootnote}{\fnsymbol{footnote}}
\begin{titlepage}
\FERMILABPub{92/27--T}
\begin{flushright}
\vspace*{-1.25in}
	FERMILAB--PUB--92/27--T\\
	ANL-HEP-PR-92-93\\
	CERN-TH 6689/92\\
	DOE-305, CPP-43\\
	FSU-HEP-920923\\
	IPNO/TH 92-91\\
	November 1992 \\
\end{flushright}
\begin{center}
{\large \bf Final-State Interaction of Longitudinal Vector Bosons}\\[.45in]
	{\bf J-L.~Basdevant}\footnotemark \\
	{\it Division de Physique Th{\'e}orique \\ Institut de Physique
Nucl{\'e}aire \\ F-91406 Orsay Cedex \\}
	{\it and} \\
	{\it LPTHE, Universit{\'e} Pierre et Marie Curie \\Paris, France}\\
[.15in]
	{\bf E.~L.~Berger} \\ {\it CERN - Geneva} \\
	{\it and} \\
	{\it High Energy Physics Division \\ Argonne
National Laboratory \\ Argonne, IL 60439} \\[.15in]
        {\bf D.~Dicus} \\ {\it Center for Particle Physics \\ University
of Texas \\Austin, TX 78712} \\[.15in]
        {\bf C.~Kao} \\
        {\it Department of Physics \\
	Florida State University \\
        Tallahassee, FL 32306} \\[.15in]
	{\bf S.~Willenbrock}\footnotemark \\
	{\it Fermi National Accelerator Laboratory \\
	P.O.~Box 500  \\
	Batavia, IL 60510} \\[1.5in]
\end{center}
\addtocounter{footnote}{-2}

\footnoterule
{\footnotemark {\footnotesize Unit{\'e} de recherche des Universit{\'e}s
Paris XI et Paris VI, associ{\'e}e au CNRS.}}

{\footnotemark {\footnotesize Permanent Address: Physics
Department, Brookhaven National Laboratory, Upton, NY 11973}}

\newpage
\end{titlepage}

\begin{document}

\renewcommand{\thefootnote}{\arabic{footnote}}
\addtocounter{footnote}{-2}

\centerline{\bf Abstract}
\bigskip

In the standard Higgs model of electroweak symmetry breaking, the
Higgs boson is associated with both vector-boson and fermion
mass generation. In contrast, we discuss a two-Higgs-doublet model
in which these masses are associated with two different scalar bosons.
We show that the Higgs boson associated with vector-boson mass
generation produces a {\it dip\/} in the cross section for $t\bar
t\to ZZ$ via a final-state interaction.  Such a dip
will be difficult to observe at the LHC/SSC.

\newpage

\leftline{\Large\bf 1.\quad Introduction}
\addtocounter{chapter}{1}
\bigskip

The most direct probe of the electroweak-symmetry-breaking mechanism is
longi\-tudinal-vector-boson scattering \cite{CG}.  For example, in the
standard Higgs model, the Higgs boson appears as a resonance in this
process \cite{CD}.

The LHC/SSC will provide the first opportunity to study
longitudinal-vector-boson scattering. There are other sources of
longitudinal-vector-boson pairs at these machines, and one might ask
if we could study longitudinal-vector-boson scattering indirectly
via the final-state interaction (rescattering) of the vector bosons.

One difficulty with posing such a question is how to
separate the effects of the final-state interaction from direct effects.
For example, in the process $gg\to V_LV_L$ ($V=W,Z$; $L$ denotes longitudinal
polarization), which proceeds via a top-quark loop, the standard-model
Higgs boson couples directly to the top quark as an $s$-channel resonance
\cite{GGMN}.
This direct effect is much larger than any
effect due to a final-state interaction.

In this paper we study a model in which the Higgs boson which appears
as a resonance in longitudinal-vector-boson scattering does not couple
directly to the top quark, so we can study its effect on $gg\to V_LV_L$
via a final-state interaction.  For simplicity we actually study
the process $t\bar t \to V_LV_L$ ($t=$ top quark), which one may regard
as a subprocess of $gg\to V_LV_L$.

We have chosen the process $gg\to V_LV_L$ because it is a large source of
longitudinal vector bosons at the LHC/SSC \cite{DKR,GV}.  The process
$q\bar q\to V_LV_L$ is also a copious source of longitudinal vector bosons
(except for $Z_LZ_L$), but almost
entirely in the $J=1$ partial wave \cite{DKR2}.  The model we study has
only $J=0$ resonances, so it does not produce large effects in the latter
process.
The process $V_TV_{T,L}\to V_LV_L$ ($T$ denotes transverse polarization)
has been
shown not to be a large source of longitudinal vector bosons \cite{BDV}.

In the next section we discuss the model we use, which is based on
two Higgs doublets.  This model is theoretically well motivated and interesting
in its own right.  In section 3 we study the process $t\bar t \to Z_LZ_L$,
including the final-state interaction of the longitudinal vector bosons.
Section 4 discusses the results and draws conclusions.
\newpage

\leftline{\Large\bf 2.\quad Two-Higgs-doublet model}
\addtocounter{chapter}{1}
\bigskip

In the standard model of the electroweak interaction, both the weak
vector bosons and the fermions acquire mass via the Higgs mechanism,
which breaks the SU(2)${}\times{}$U(1)  symmetry down to
electromagnetism. The symmetry is broken by an SU(2) scalar doublet
which acquires a vacuum-expectation value. A scalar particle, dubbed
the Higgs boson, becomes part of the physical spectrum. Thus the Higgs
boson is associated with both vector-boson and fermion mass generation.

In the absence of the Higgs boson or, more generally, a model for the
symmetry-breaking mechanism, the tree amplitude for
longitudinal-vector-boson scattering is proportional to $g^2s/M^2_W$,
and violates the unitarity bound at
$s \approx 4\pi\sqrt 2/G_F \approx (1.2\; {\rm TeV})^2$ \cite{CG,MVW}.  The
Higgs boson, exchanged in the $s$, $t$, and $u$ channels, cancels the
terms in the amplitude which grow with energy, leaving an amplitude
proportional to $g^2m^2_H/M^2_W$ \cite{DM,LQT}. Thus we may say that
the Higgs boson is responsible for ``unitarizing'' the
longitudinal-vector-boson scattering amplitude, in the sense that the
unitarity bound is respected at all energies.

Similarly, the amplitude for $t\bar t\to V_LV_L$,
with the $t$ and $\bar t$ of
the same helicity, is proportional to $g^2
m_t\sqrt{s}/M^2_W$ \cite{AC}. The unitarity bound is violated at an
energy which depends on $m_t$ \cite{AC,MVW}
\footnote{The unitarity bound of Ref.~\cite{AC} is considerably strengthened
in Ref.~\cite{MVW} by considering the $I=0$, $J=0$, spin zero, color-singlet
amplitude.},
\begin{equation}
\sqrt{s} \approx \frac{4\pi\sqrt{2}}{3G_Fm_t}. \label{eq:crit}
\end{equation}
The Higgs boson, exchanged in the $s$ channel, again cancels the terms
which grow with energy, leaving terms proportional to $g^2
(m^2_H/M^2_W) m_t/\sqrt{s}$ and $g^2 (m^2_t/M^2_W)
m_t/\sqrt{s}$ \cite{CFH}. Thus the Higgs boson also
``unitarizes'' $t\bar t\to V_LV_L$.

One may regard the standard Higgs model as economical, in that it
generates both vector-boson and fermion masses via the same mechanism.
However, we have no guarantee that nature has chosen this model, so we
should keep an open mind regarding the possible manifestations of the
vector-boson and fermion mass-generating mechanisms. These two
mechanisms could be quite distinct and have very different
experimental signatures. This point has been appreciated at least
since the development of technicolor \cite{TECH} and extended
technicolor \cite{EXT}, and it has
recently been emphasized in Ref.~\cite{BC}. Note, however, that in
technicolor/extended technicolor both $V_LV_L \to V_LV_L$ and
$t\bar t \to V_LV_L$ are unitarized by the same condensate, at a
scale ${\cal O}(4\pi\sqrt 2/G_F)$ \cite{AC}.

In this paper we consider what is possibly the simplest model in which
the vector-boson and fermion masses are generated by separate
mechanisms: a two-Higgs-doublet model. One may introduce a discrete
symmetry to ensure that only one doublet couples to fermions \cite{GW}.
If this doublet has a small vacuum-expectation value,
it couples only weakly to vector bosons, but with correspondingly
enhanced strength to
fermions. The neutral scalar boson associated with this doublet
unitarizes $t\bar t\to V_LV_L$, but it contributes very little to the
unitarization of $V_LV_L\to V_LV_L$, which is unitarized almost
entirely by the neutral scalar boson associated with the other
doublet. This model has been considered previously in Ref.~\cite{HKS}
with a rather different motivation.  We do not regard this model as a
candidate for a fundamental theory, but as the simplest model which embodies
the philosophy described above, and useful for suggesting
signatures for the vector-boson and fermion mass-generating mechanisms.

\bigskip\bigskip
\leftline{\Large\bf 3.\quad $t\bar t \to Z_LZ_L$}
\addtocounter{chapter}{1}
\addtocounter{equation}{-1}
\bigskip
For calculational expediency, we consider only terms of enhanced
electroweak strength, ${\cal O}(g^2m^2/M^2_W)$, where $m$ is a
Higgs-boson or top-quark mass. Formally, this choice corresponds to the
limit $g\to0$ with $v$ fixed ($M_W=\frac{1}{2} gv$). The polarization
vectors of the longitudinal vector bosons can be approximated by
$\epsilon^\mu_L(p) \approx p^\mu/M_V$ in this limit.

The tree-level Feynman diagrams for $t\bar t\to Z_LZ_L$ are shown in
Fig.~1. The necessary Feynman rules can be found in Refs.~\cite{HKS,GH}
\footnote{Although the appendix
of Ref.~\cite{GH} specializes to the supersymmetric
two-Higgs-doublet model, the $HVV$ and
$Ht\bar t$ couplings are valid
for the model considered here, with $H^0=H_1$, $h^0=H_2$.}.
The amplitude, with the $t$ and $\bar t$ of the same helicity and
opposite color, is
\begin{eqnarray}
A_0 = &-& \frac{g^2m_t}{4M^2_W} \bar v(p_2) \left[1 + m_t \left(
\frac{1}{\not\! p_3 - \not\! p_1 -m_t} + \frac{1}{\not\! p_2-\not\! p_3
-m_t}\right) \right. \label{eq:tree} \\
&-& \left.\frac{1}{\sin \beta} \left( \sin\alpha \cos(\beta-\alpha)
\frac{s}{s-m^2_1+im_1\Gamma_1} + \cos \alpha\sin (\beta-\alpha)
\frac{s}{s-m^2_2+im_2\Gamma_2} \right) \right]u(p_1) \nonumber
\end{eqnarray}
where $m_t$ is the top-quark mass and $m_{1,2}$ are the masses of the
Higgs bosons. The parameter $\beta$ is related to the ratio of the
vacuum-expectation values of the two Higgs doublets by
$\tan\beta=v_2/v_1$; we consider $\tan\beta<<1$. The
parameter $\alpha$ mixes the neutral scalar Higgs bosons from the two
doublets.  The case under consideration corresponds to no mixing, so we
set
$\alpha=0$. One then sees that the
$H_1$-exchange diagram vanishes, and the $H_2$-exchange diagram alone is
responsible for ``unitarizing'' the amplitude, i.e., canceling the
first term in Eq.~(\ref{eq:tree}) at high energy.

We now wish to calculate the effect of a final-state interaction
(rescattering) of the longitudinal vector bosons. Away from the $H_1$
resonance, the longitudinal vector bosons are weakly coupled, and the effect
of rescattering is small. We therefore need only consider diagrams
involving an $s$-channel $H_1$ boson. The contributing Feynman
diagrams, in 't~Hooft-Feynman gauge, are shown in Fig.~2. An $H_2H_1$
counterterm is necessary to absorb the ultraviolet divergence in the
second diagram. This counterterm renormalizes the mixing parameter
$\alpha$ from its bare value \cite{HKS}.
We set the renormalized $\alpha$ to
zero. The necessity to retune the renormalized $\alpha$ to zero at one
loop is a consequence of the fact that $\alpha=0$ (or any other value)
is not enforced by any symmetry in this model.  In effect we are taking
the short-distance part of the loop integral, canceling it against the
bare $\alpha$, and regarding the remaining finite part as a long-distance
final-state interaction.

The one-loop amplitude for $t\bar t\to Z_LZ_L$ is
\footnote{The coupling of the Goldstone bosons to the Higgs bosons is
obtained by multiplying the corresponding $H_iVV$ coupling by
$-m_i^2/2M_V^2$ and dropping the $g^{\mu\nu}$. Note that the couplings in
the appendix of
Ref.~\cite{GH} pertain specifically to the supersymmetric two-Higgs-doublet
model.}
\begin{eqnarray}
A_1 &=& \quad-\;\frac{1}{(4\pi)^2}\;\frac{g^4m^2_1m_t}{16M^4_W}
\cos^2(\beta-\alpha) \frac{s}{s-m^2_1+im_1\Gamma_1}\\ \nonumber
& &\quad\bar v(p_2) \biggl[m_t^2\left(C^Z_{11}-C^Z_{12}+(1-\gamma_5)(C^W_{11}
+ C^W_0) -(1+\gamma_5) C^W_{12}\right) \\ \nonumber
& &\quad + \frac{\cos\alpha\sin(\beta-\alpha)}{\sin\beta} \;
\frac{m^2_2}{s -m^2_2+im_2\Gamma_2} \left( B^W_0
+\frac{1}{2}B^Z_0\right)\biggr] u(p_1) \label{eq:loop}
\end{eqnarray}
where
\begin{eqnarray*}
&\frac{1}{i\pi^2}\int d^4k &
\frac{k^\mu}{[k^2-M_Z^2][(k+p_1)^2-m_t^2][(k+p_1+p_2)^2-M^2_Z]} \\
& & = C^Z_{11}p^\mu_1 + C^Z_{12}p^\mu_2 \\
&\frac{1}{i\pi^2}\int d^4k & \frac{\{1;k^\mu\}}{[k^2-M_W^2] [(k+p_1)^2-m_b^2]
[(k+p_1+p_2)^2-M^2_W]} \\
& & = \{C^W_0;\; C^W_{11}p^\mu_1 + C^W_{12}p^\mu_2\}
\end{eqnarray*}
and
\begin{eqnarray*}
B^V_0 & = & \frac{1}{i\pi^2} \int d^4k \frac{1}{[k^2-M^2_V]
[(k+p_1+p_2)^2 - M^2_V]} + {\rm counterterm}\\
& = & -(1-4M^2_V/s)^{1/2}\ln\left(
\frac{(1-4M^2_V/s)^{1/2}+1}{(1-4M^2_V/s)^{1/2}-1}\right) +2\;.
\end{eqnarray*}
The counterterm has been chosen such
that $B^V_0=0$ at $s=0$. Note that $B^V_0$ has a (positive) imaginary part for
$s>4M^2_V$. The loop integrals were evaluated with the codes FF
\cite{FF} and LOOP \cite{LOOP}
\footnote{LOOP is a completely FORTRAN version of
Veltman's FORMFactor \cite{FORMF}.}.

Although any value of $\beta$ is theoretically acceptable, a very
small value enhances the top-quark Yukawa coupling such that it
becomes strong, and perturbation theory breaks down.  Unitarity of
$t\bar t \to t\bar t$ suggests that this occurs for \cite{CFH,MVW}
\begin{equation}
\sin^2\beta \approx \frac{3G_Fm_t^2}{4\pi\sqrt{2}}\cos^2\alpha. \label{eq:beta}
\end{equation}
Since we are interested in performing a perturbative calculation, we
choose a value of $\beta$ greater than that given by Eq.~(\ref{eq:beta}).

The two-Higgs-doublet model also contains a charged scalar, $H^{\pm}$,
and a neutral pseudoscalar, $A$.  These particles, as well as $H_2$,
contribute to $H_1$ production
at one loop via diagrams in which they replace one or both of
the Goldstone bosons in Fig.~2.
The cubic scalar interactions are model dependent, so we have
not included these contributions.  We argue in the next section that
they do not change the qualitative
results of our calculation.

\bigskip\bigskip
\leftline{\Large\bf 4.\quad Results and conclusions}
\addtocounter{chapter}{1}
\addtocounter{equation}{-3}
\bigskip
As a typical example of what one would expect to observe,
we show in Fig.~3 the square of the zeroth partial wave of $t\bar t\to
Z_LZ_L$, with the $t$ and $\bar t$ of the same helicity and opposite
color,
for $m_1=500$ GeV, $m_2=300$ GeV, $m_t=130$ GeV, $\beta=0.3$, and $\alpha=0$.
The $H_2$ resonance, which couples directly to the
top quark, produces the expected peak. The $H_1$, which does not couple
directly to the top quark, produces a {\it dip}. Such dips are
generic for a final-state interaction which proceeds via a resonance,
and are known in hadronic
physics \cite{R,BB,MP}.  This phenomenon is reviewed in an appendix,
both diagrammatically
and via the Omn\`es-Muskhelishvili formalism \cite{MO,J}.
A simple understanding of this phenomenon
is gained by noting that the absorptive
part of the loop diagrams, obtained via the Cutkosky rules by ``cutting''
the Goldstone-boson propagators, is proportional to the product of the
tree amplitude $t\bar t \to V_LV_L$ and the vector-boson-scattering
amplitude $V_LV_L \to V_LV_L$ (using the equivalence of the Goldstone
bosons and $V_L$ in the limit $g\to 0$).  At the peak of the $H_1$
resonance, the latter amplitude is purely imaginary; thus the absorptive
part of the loop diagrams interferes destructively with the tree
amplitude, producing a dip.  Including $H_2, H^{\pm}$, and $A$ in the loop
diagrams does not change this result qualitatively, as is made clear in
the appendix.

Unfortunately, a dip in the $V_LV_L$ invariant mass spectrum will be very
difficult to observe at the LHC/SSC. There is a large continuum
background from $q\bar q \to V_TV_T$ \cite{DKR2} and $gg\to V_TV_T$
\cite{DKR,GV} which is typically an order of magnitude larger than the
$gg\to V_LV_L$ continuum.  At the $H_2$ resonance, $gg\to V_LV_L$
is enhanced such that the signal is comparable to or greater than the
background.  However, a dip in the $gg\to V_LV_L$ process near the $H_1$
mass will be difficult to distinguish on the large continuum background.
Furthermore,
the $H_1$ will appear as a resonance in longitudinal-vector-boson scattering,
so one would have to separate this process from $gg\to V_LV_L$ in order
to have any chance of observing the dip in the latter.  Such a separation
may be possible by tagging the forward jets associated with
longitudinal-vector-boson scattering, but it is not one-hundred percent
efficient \cite{JETTAG}.

We conclude that
while $t\bar t \to V_LV_L$ (via $gg\to V_LV_L$ at the LHC/SSC)
is likely to tell us about the mass-generating mechanism for the top quark,
it is unlikely to reveal information on the mass-generating
mechanism for the vector bosons via the final-state interaction of the
longitudinal vector bosons.
If the particles associated with vector-boson mass generation couple
to top quarks (as in the standard Higgs model) or to gluons \cite{BDV2},
they could manifest themselves directly in the process
$gg\to V_LV_L$.

\newpage
\leftline{\bf Acknowledgements}
\medskip

\noindent We are grateful for conversations with
F.~Boudjema, H.~Haber, W.~Marciano,
F.~Paige, M.~Pennington, C.~Quigg, V.~Teplitz,
L.~Trueman, and G.~Valencia.
J-L.~B. is grateful for the kind hospitality extended to him by the High
Energy Physics Division of Argonne National Laboratory where part of this
work was performed.
E.~L.~B. was supported in part by the U.~S. Department of Energy, Division
of High Energy Physics, Contract W-31-109-ENG-38.
D.~D. was supported in part by
the U.~S. Department of Energy under contract number DE-FG05-ER8540200.
C.~K. was supported in part by
the U.~S. Department of Energy and in part by the Texas National Research
Laboratory Commission.
S.~W. was supported by an award from
the Texas National Research Laboratory Commission and by contract number
DE-AC02-76 CH 00016 with the U.~S.~Department of Energy.
\newpage

\appendix
\chapter{}

We show that a final-state interaction which proceeds via a resonance
produces a dip near the resonance mass. The argument we give is a
generalization of that given in the appendix of the first paper of
Ref.~\cite{BB}. We
also show that the same result can be obtained via the
Omn\`es-Muskhelishvili formalism \cite{MO,J}.

If we denote by $g$ the coupling of the vector bosons, of mass $M$, to
the spin-zero resonance, the zeroth partial wave of the elastic
scattering amplitude of the vector bosons is
\begin{equation}
\frac{1}{\beta} e^{i\delta}\sin\delta = -\frac{g^2}{16\pi}\;
\frac{1}{s-m^2+ g^2 \Pi} \label{eq:aone}
\end{equation}
where $\beta=(1-4M^2/s)^{1/2}$, $\delta$ is the phase shift, and
$\Pi=B^V_0/(4\pi)^2$ is the renormalized two-point function.
Let us denote by $a_B$ the zeroth partial wave of the weak production
amplitude of the strongly-interacting vector bosons.
The zeroth partial wave of the full amplitude, given by the sum of
$a_B$ and the one-loop amplitude formed by the
rescattering of the vector bosons through the resonance (see Fig.~4), is
\begin{eqnarray}
a & =  & a_B - g^2 L \frac{1}{s-m^2+ g^2\Pi}
\nonumber \\
 & = & \frac{1}{s-m^2 + g^2\Pi}\; \left[a_B \left(s-m^2 +
g^2\Pi\right) - g^2 L\right] \label{eq:atwo}
\end{eqnarray}
where $L$ is the loop integral associated with the loop diagram.  Unitarity,
via the Cutkosky rules, tells us
\begin{equation}
\mbox{Im}\; \Pi = \frac{1}{16\pi}\beta \label{eq:athree}
\end{equation}
and
\begin{equation}
\mbox{Im}\; L = \frac{1}{16\pi}\beta a_B \label{eq:afour}
\end{equation}
so the numerator of the amplitude is real, with a zero near $s=m^2$.
The real parts of the loop integrals shift the position of the zero.

The Omn\`es-Muskhelishvili formalism provides a means of implementing unitarity
and analyticity \cite{MO,J}. Writing and solving a dispersion relation for
$a-a_B$, one obtains \cite{R,BB,MP,MO,J}
\begin{equation}
a=a_B-\Omega \frac{1}{\pi} \int^{\infty}_{4M^2}
\frac{ds^{\prime}}{s^{\prime}-s}
[\mbox{Im}\; \Omega^{-1}]a_B \label{eq:afive}
\end{equation}
where
\begin{eqnarray}
\Omega & = & \exp \left[\frac{s}{\pi} \int^{\infty}_{4M^2}
\frac{ds^{\prime}}{(s^{\prime}-s)s^{\prime}} \delta (s^{\prime}) \right]
\nonumber \\
& = & - \frac{m^2}{s-m^2 + g^2\Pi} \label{eq:asix}
\end{eqnarray}
is the Omn\`es function for the phase shift given by Eq.~(\ref{eq:aone})
\cite{BB}.
Eq.~(\ref{eq:afive}) may thus be written
\begin{equation}
a=a_B - g^2 \; \frac{1}{s-m^2 + g^2\Pi}
\left[\frac{1}{\pi} \int^{\infty}_{4M^2}
\frac{ds^{\prime}}{s^{\prime}-s} \beta a_B\right].
\label{eq:aseven}
\end{equation}
The factor in the bracket is the integral representation of
$L$, which proves the equivalence of Eqs.~(\ref{eq:aseven}) and
(\ref{eq:atwo}). A counterterm, or subtraction constant, is necessary
if $L$ is ultraviolet divergent.

One may add an additional term, $P\Omega$, to Eq.~(\ref{eq:afive}),
where $P$ is an arbitrary
real polynomial, and still satisfy the unitarity and analyticity
requirements of the Omn\`es-Muskhelishvili formalism. For
$P={}$constant, this term represents a direct coupling of the
resonance to the initial state.
In Ref.~\cite{P} the effect of a final-state interaction via a
technirho resonance on $e^+e^-\to W^+W^-$ is represented by
$a_B\Omega$, resulting in a resonant enhancement near the technirho
mass. Our arguments show that this term corresponds to a direct
coupling of the technirho, not to a final-state interaction. In
Ref.~\cite{I} the technirho is introduced via the Gounaris-Sakurai
model, which is based on vector-meson dominance. This also corresponds
to a direct coupling, not to a final-state interaction, as claimed in
that work.

The phase variation of $e^+e^-\to W^+W^-$ due to a final-state interaction
is discussed in Ref.~\cite{HI}.

\newpage
{\bf Figure captions}\\

Fig.~1 - Feynman diagrams for $t\bar t \to Z_LZ_L$ in a two-Higgs-doublet
model. For $\alpha=0$, only $H_2$ contributes to the last diagram.
\bigskip

Fig.~2 - Feynman diagrams of enhanced electroweak strength for
$t\bar t \to Z_LZ_L$ near the $H_1$ resonance, in 't~Hooft-Feynman gauge.
The Goldstone bosons are denoted by $w^{\pm},z$.  One may regard these
diagrams as representing a final-state interaction of the longitudinal
vector bosons via the $H_1$ resonance.
\bigskip

Fig.~3 - Square of the zeroth partial wave of  $t\bar t \to Z_LZ_L$, with
the $t$ and $\bar t$ of the same helicity and opposite color, versus
center-of-mass energy, for $m_1=500$ GeV, $m_2=300$ GeV, $m_t=130$ GeV,
$\beta=0.3$, and $\alpha=0$.  Both the tree amplitude (dashed) and the
amplitude including the final-state interaction of the longitudinal vector
bosons (solid) are shown.  The $H_2$ resonance produces the expected peak,
while the $H_1$, which does not couple directly to the top quark,
produces a dip.
\bigskip

Fig.~4 - Weak production amplitude plus the one-loop amplitude formed by the
rescattering of the final particles through a resonance.


\newpage
\begin{thebibliography}{999}
\bibitem{CG} M.~Chanowitz and M.~K.~Gaillard, Nucl.~Phys.~{\bf B261},
379 (1985).
\bibitem{CD} R.~Cahn and S.~Dawson, Phys.~Lett.~{\bf 136B}, 196 (1984);
G.~Kane, W.~Repko, and W.~Rolnick, Phys.~Lett.~{\bf 148B}, 367 (1984).
\bibitem{GGMN} H.~Georgi, S.~Glashow, M.~Machacek, and D.~Nanopoulos,
Phys.~Rev.~Lett.~{\bf 40}, 692 (1978).
\bibitem{DKR} D.~Dicus, C.~Kao, and W.~Repko, Phys.~Rev.~D~{\bf 36}, 1570
(1987); D.~Dicus and C.~Kao, Phys.~Rev.~D~{\bf 43}, 1555 (1991).
\bibitem{GV} E.~W.~N.~Glover and J.~van der Bij, Nucl.~Phys.~{\bf B321}, 561
(1989).
\bibitem{DKR2} M.~Duncan, G.~Kane, and W.~Repko, Nucl.~Phys.~{\bf B272},
517 (1986); M.~Duncan, Phys.~Lett.~{\bf 179B}, 393 (1986).
\bibitem{BDV} J.~Bagger, S.~Dawson, and G.~Valencia, Fermilab-Pub-92/75-T.
\bibitem{MVW} W.~Marciano, G.~Valencia, and S.~Willenbrock,
Phys.~Rev.~D~{\bf 40}, 1725 (1989).
\bibitem{DM} D.~Dicus and V.~Mathur, Phys.~Rev.~D~{\bf 7}, 3111 (1973).
\bibitem{LQT} B.~Lee, C.~Quigg, and H.~Thacker, Phys.~Rev.~Lett.~{\bf 38},
883 (1977); Phys.~Rev.~D~{\bf 16}, 1519 (1977).
\bibitem{AC} T.~Appelquist and M.~Chanowitz, Phys.~Rev.~Lett.~{\bf 59}, 2405
(1987).
\bibitem{CFH} M.~Chanowitz, M.~Furman, and I.~Hinchliffe,
Phys.~Lett.~{\bf 78B}, 285 (1978); Nucl.~Phys.~{\bf B153}, 402 (1979).
\bibitem{TECH} S.~Weinberg, Phys.~Rev.~D~{\bf 19}, 1277 (1979);
L.~Susskind, Phys.~Rev.~D~{\bf 20}, 2619 (1979).
\bibitem{EXT} S.~Dimopoulos and L.~Susskind, Nucl.~Phys.~{\bf B155}, 237
(1979);
E.~Eichten and K.~Lane, Phys.~Lett.~{\bf 90B}, 125 (1980).
\bibitem{BC} M.~Berger and M.~Chanowitz, Phys.~Rev.~Lett.~{\bf 68}, 757 (1992).
\bibitem{GW} S.~Glashow and S.~Weinberg, Phys.~Rev.~D~{\bf 15}, 1958 (1977).
\bibitem{HKS} H.~Haber, G.~Kane, and T.~Sterling, Nucl.~Phys.~{\bf B161}, 493
(1979).
\bibitem{GH} J.~Gunion and H.~Haber, Nucl.~Phys.~{\bf B272}, 1 (1986);
J.~Gunion, H.~Haber, G.~Kane, and S.~Dawson, {\it The Higgs Hunter's Guide}
(Addison-Wesley, New York, 1990).
\bibitem{FF} G.~J.~van Oldenborgh, Comp.~Phys.~Comm.~{\bf 66}, 1 (1991).
\bibitem{LOOP} D.~Dicus and C.~Kao, LOOP, a FORTRAN program for doing loop
integrals of 1, 2, 3, and 4 point functions with momenta in the numerator,
1991, unpublished.
\bibitem{FORMF} G.~Passarino and M.~Veltman, Nucl.~Phys.~{\bf B160}, 151
(1979).
\bibitem{H} H.~Haber, in {\it Proceedings of the International Workshop on
Electroweak Symmetry Breaking}, Hiroshima, Japan (1991).
\bibitem{R} L.~Resnick, Phys.~Rev.~D~{\bf 2}, 1975 (1970); J.~Pumplin, Phys.~
Rev.~D~{\bf 2}, 1859 (1970); T.~Bauer, Phys.~Rev.~Lett.~{\bf 25}, 485 (1970).
\bibitem{BB} J-L.~Basdevant and E.~Berger, Phys.~Rev.~D~{\bf 16}, 657 (1977);
D~{\bf 19}, 239 (1979); D~{\bf 19}, 246 (1979).
\bibitem{MP} D.~Morgan and M.~Pennington, Z.~Phys.~{\bf C37}, 431 (1988).
\bibitem{MO} R.~Omn{\`e}s, Il Nuovo Cim. {\bf 8}, 316 (1958);
N.~Muskhelishvili, {\it Singular Integral Equations} (Groningen, 1953).
\bibitem{J} J.~D.~Jackson, in {\it Dispersion Relations}, Scottish
Universities'
Summer School, 1960, ed. G.~Screaton (Oliver and Boyd, Edinburgh, 1961), p.~1.
\bibitem{JETTAG} R.~Cahn, S.~Ellis, R.~Kleiss, and W.~J.~Stirling, Phys.~
Rev.~{\bf D35}, 1626 (1987).
\bibitem{BDV2} J.~Bagger, S.~Dawson, and G.~Valencia, Phys.~Rev.~Lett.~
{\bf 67}, 2256 (1991); Phys.~Lett.~{\bf 292B}, 137 (1992).
\bibitem{P} M.~Peskin, in {\it Physics in Collision IV}, ed. A.~Seiden
(\'Editions Fronti\`eres, Gif-sur-Yvette, 1984); talk presented at
the International Workshop on Physics and Experiments with Linear Colliders,
Saariselka, Finland, Sept.~9-14, 1991, SLAC-PUB-5798 (1992).
\bibitem{I} F.~Iddir, A.~Le~Yaouanc, L.~Olivier, O.~P\`ene, and J.-C.~Raynal,
Phys.~Rev.~D~{\bf 41}, 22 (1990).
\bibitem{HI} K.~Hikasa, talk presented at
the International Workshop on Physics and Experiments with Linear Colliders,
Saariselka, Finland, Sept.~9-14, 1991, KEK-TH-319 (1992).
\end{thebibliography}
\end{document}